\documentstyle[preprint,aps,epsf]{revtex}

\begin{document}
\title{Visualization of Coulomb Correlations in Finite Metallic Systems}
\author{F. Despa and R.S. Berry}
\address{Department of Chemistry,\\
The University of Chicago,\\
Chicago, Illinois 60637}
\maketitle

\begin{abstract}
We present an analytic ansatz to find the effective electrostatic potential
and Coulomb correlations in multicenter problems, specifically homogeneous
and doped clusters of metal atoms. The approach is based on a
quasi-classical density-functional treatment. We focus on the interpretive
aspect of our findings, particularly on extracting insight regarding the
geometric effects of Coulomb correlations for any given spatial disposition
of ionic cores. For singly-doped metallic clusters we obtain a direct
visualization of the variations of both screening and Coulomb correlations
with changes of location of the dopant atom. This analysis provides a way to
interpret recent observations of the variability of physical properties of
metal clusters with changes of composition and geometry.
\end{abstract}

\newpage

Collective effects induced by Coulomb correlations in atoms have been
studied in two ways. In the first, both hydrodynamic theory and local
approximate dielectric theory have been used; neither of these takes into
account either shell structure or the single-particle spectrum of the
valence electrons.\cite{lundqvist} These methods are capable, at most, of
giving gross trends in dynamical properties. The second route instead uses a
fully quantal description based on the one-electron excitation spectrum and
corresponding wavefunctions. A recent collection of papers provides a
description of methods and results of the application of many-body
techniques in atomic theory.\cite{lindgren}

The way electrons are correlated can be inferred from the probability
distribution implied by their wavefunction. To make it possible to make such
inferences, however, we must be a bit thoughtful about how we present this
distribution. Even for a two-electron atom, we begin with a function of six
independent variables in a fixed center-of-mass system. We would like to
extract from this a description in no more than two or three independent
variables, something we can represent pictorially and visualize. For a
three-body system such as $He^{**}$ or the valence electrons of $Mg$, a
natural and practical way to carry out such a description has emerged as an
analytic reduction of the probability density $\left| \Psi \left({\bf %
r_{1},r_{2}}\right) \right| ^{2}$ to the joint probability density $p\left(
r_{1},r_{2},\theta _{12}\right)$, where $\theta _{12}$ is the angle between
the position vectors $r_{1}$ and $r_{2}$ of two electrons.\cite{berry,berry1}
This in turn makes it straightforward to compute and display the conditional
probability density $d\left( r_{2},\theta _{12};r_{1}\right)$, for finding
one of the two electrons at distance $r_1$ from the nucleus, and at an angle 
$\theta_{12}$ from the vector from the nucleus to electron $2$ if electron $%
2 $ is at a distance $r_2$ from the nucleus. Three-dimensional graphs of $%
d\left(r_{1},r_{2},\theta _{12}\right) $ provide a vivid and precise way to
depict the correlation of two electrons.\cite{berry0} With this probability
density $d$, one can compare wave functions of different qualities, see what
roles long-range and short range correlation play in various states, exhibit
the relative importance of angular and radial correlation, and compare
correlation in different atomic systems. The work, started in the late $%
1970s $ by Rehmus, Kellman, Roothaan and Berry\cite{berry,berry1}, provides
a generalization of other quantitative descriptions of electron correlation.%
\cite
{chandrasekhar,coulson,dickens,sinanoglu,wulfman,herrick,banyard,fano,banyard1}

For the system of more particles, we have yet to find a comparably powerful
approach because so much information is contained in the wave function and
we do not know how to extract what is relevant in a manner adaptable to
pictures. An exception emerges, obviously, for the high-density limit of the
Fermi fluid where a collective description of electrons is likely to be
optimal. This collective description is based on the organized behavior of
the electrons brought about by their long-range Coulomb interactions. The
long-range Coulomb interactions, subject to the screening among the electric
charges, act to couple together the motion of many electrons, giving rise to
the well-known quantum density oscillations. \cite{friedel}

Working toward extracting insights from the probability density $\left| \Psi
\right| ^{2}$ and making use of the quasi-classical description for the
(valence) electron gas, we develop here an analytical ansatz which allows us
to find and visualize the effective electrostatic potential and Coulomb
correlations in multicenter problems. We apply this ansatz to the case of
moderately large metallic clusters. To anticipate what follows, let us state
our findings: by using a generalized partition function for valence
electrons (the Bloch density matrix), the electron self-distribution in the
common potential $V\left( {\bf r}\right) $ is derived in terms of many-body
perturbation theory.\cite{march1} This approach produces the electron
density $\rho \left( {\bf r}\right)$ as a functional of $V\left({\bf r}%
\right)$ (with $\rho\left( {\bf r}\right)\sim V\left({\bf r}\right) $),
which is valid for describing metallic systems, i.e. systems with a
high-density valence electron gas. Further, inside the electron gas of
density $\rho \left( {\bf r}\right) $, we introduce the cluster cage formed
by the positive ion cores with the spatial distribution given by $\rho
_{+}\left( {\bf R}_{i}\right) $, and apply Poisson's equation to the cluster
as a whole. (The vectors ${\bf R}_{i}$ are position vectors of the ions.)
The self-consistent solution of this equation gives the collective
description of the cluster constituents, electrons plus ions. This
generalization of the Coulomb interaction results in a superposition of
quantum oscillations given by long-range contributions and screening on the
smooth ''semiclassical'' potential \cite{note1}. We focus on their
interpretive aspects and specifically on extracting insights regarding the
geometric effects of Coulomb correlations for any given spatial disposition
of ionic cores. Also, we explore the case of a foreign metal atom doping a
otherwise-homogeneous cluster of metal atoms. The approach presented here
provides us with a direct visualization of the way both the screening effect
and the Coulomb correlations change with changes of the location of the
impurity. This analysis is important in the context of recent observations
of the role played by composition and geometry in changing the physical
properties of metallic clusters \cite{heinebrodt,class,akola}.

Consider a fixed positive ion distribution in space $\rho _{+}\left( {\bf R}%
_{i}\right) $, with ${\bf R}_{i}$ the positions of the ions measured from
the center of the cluster. The ion cluster cage has a net charge measured in
appropriate units equal to $zN$, where $z$ is the electric charge of one ion
(for simplicity, we shall restrict the discussion to single-valent metals, $%
z=1$) and $N$ denotes the total number of ions in the cluster. If a gas of
valence electrons carrying an equal number of negative charges is
introduced, so that the system is strictly neutral, then the electrons
redistribute themselves so as to shield the positive charges at large
distances and minimize the Coulomb self-energy of that gas, and also satisfy
the Fermion constraints on the electrons. In the high density limit of the
Fermi fluid, Bohr's Correspondence Principle applies and we can introduce
the collective description of the electron gas based on the methods of
statistical mechanics. The generalized partition function of the valence
electrons moving in the common potential $V\left( {\bf r}\right) $ can be
written in terms of the wave functions $\Psi _{i}\left( {\bf r}\right) $ and
energy levels $\varepsilon _{i}$ as 
\begin{equation}
\Gamma \left( {\bf r}^{\prime },{\bf r,}\beta \right) =\sum_{i}\Psi
_{i}^{*}\left( {\bf r}^{\prime }\right) \Psi _{i}\left( {\bf r}\right)
e^{-\varepsilon _{i}\beta }\;\;,  \label{one}
\end{equation}
where $\beta =\left( k_{B}T\right) ^{-1}$, $k_{B}$ is Boltzmann's constant
and $T$ the absolute temperature. By integrating this along the diagonal,
where ${\bf r}^{\prime }={\bf r}$, we obtain the ordinary partition function
of statistical mechanics. Eq. $\left( 1\right) $ is the Bloch density matrix
and if we operate with the one-particle Hamiltonian 
\begin{equation}
H_{s}=-\frac{1}{2}\nabla _{{\bf r}}^{2}+V\left( {\bf r}\right) \;\;,
\label{two}
\end{equation}
on $\Gamma $, and compare the result with that obtained by differentiating $%
\Gamma $ with respect to $\beta $, then we find the Bloch equation 
\begin{equation}
H_{s}\Gamma =-\frac{\partial \Gamma }{\partial \beta }\;\;,  \label{three}
\end{equation}
which has the form of the time-dependent Schr\"{o}dinger eq., with $\beta $
playing the role of $it$. The boundary condition required to define the
solution of $\left( 3\right) $ follows from the completeness theorem for
eigenfunctions, namely 
\[
\Gamma \left( {\bf r}^{\prime },{\bf r,}0\right) =\delta \left( {\bf r}%
^{\prime }-{\bf r}\right) \;\;. 
\]
In the high density limit, the behavior of the electrons is simple, and the
Coulomb interaction can be treated as a perturbation of the motion of the
free electrons. Therefore the solution of eq. $\left( 3\right) $ may be
written as 
\begin{eqnarray}
\Gamma \left( {\bf r}^{\prime },{\bf r,}\beta \right) &=&\Gamma _{0}\left( 
{\bf r}^{\prime },{\bf r,}\beta \right) -  \label{four} \\
&&\int d{\bf r}^{\prime \prime }\int_{0}^{\beta }d\beta ^{\prime }\Gamma
_{0}\left( {\bf r}^{\prime },{\bf r}^{\prime \prime }{\bf ,}\beta {\bf -}%
\beta ^{\prime }\right) V\left( {\bf r}^{\prime \prime }\right) \Gamma
\left( {\bf r}^{\prime \prime },{\bf r,}\beta ^{\prime }\right) \;\;, 
\nonumber
\end{eqnarray}
where $\Gamma _{0}$ is the Bloch density matrix for an assembly of free
electrons. The derivation of Dirac's density matrix $\gamma \left( {\bf r}%
^{\prime },{\bf r}\right) $ from $\left( 4\right) $ is described in Ref. 
\cite{march1}. It consists of an iterative procedure to find the
perturbation terms in $\Gamma $. The first step is replacement of $\Gamma
_{0}$ for $\Gamma $ in the integral and integration over $\beta ^{\prime }$.
If the Bloch density matrix $\Gamma $ is determined, Dirac's density matrix $%
\gamma $ (which is actually the electron density) may be obtained by using
the Laplace transform relation connecting $\Gamma \left( {\bf r}^{\prime },%
{\bf r,}\beta \right) $ and $\gamma \left( {\bf r}^{\prime },{\bf r,}\xi
\right) $%
\[
\Gamma \left( {\bf r}^{\prime },{\bf r,}\beta \right) =\beta
\int_{0}^{\infty }d\xi \gamma \left( {\bf r}^{\prime },{\bf r,}\xi \right)
\exp \left( -\beta \xi \right) \;\;, 
\]
where $\xi $ is an intensive energy variable conjugate to $\beta $.

Following the above procedure, the electron density is obtained as 
\begin{equation}
\rho \left( {\bf r}\right) =\rho _{0}-\frac{k_{F}^{2}}{2\pi ^{3}}\int d{\bf r%
}^{\prime }\;V\left( {\bf r}^{\prime }\right) \frac{j_{1}\left( 2k_{F}\left| 
{\bf r-r}^{\prime }\right| \right) }{\left| {\bf r-r}^{\prime }\right| ^{2}}%
\;\;\;.  \label{five}
\end{equation}
Here $k_{F}$ is the Fermi wavevector, $\rho _{0}$ is the free-particle
density, $\rho _{0}=\frac{k_{F}^{3}}{3\pi ^{2}}$, and $j_{1}\left( x\right) $
is the first-order spherical Bessel function.

With the primary form $\left( 5\right) ,$ the electron density $\rho \left( 
{\bf r}\right) $ makes further mathematical computations very difficult. To
go further, we need to simplify by linearizing. Usually, this linearization
proceeds by adding the assumption that $V\left( {\bf r}\right) $ varies
slowly in space, the Thomas-Fermi approximation. Accordingly, $V\left({\bf r}%
^{\prime }\right) $ is replaced in $\left( 5\right) $ by $V\left( {\bf r}
\right) $ \cite{mott}. One obvious point needs to be stressed here: the
discrete positive ion distribution used here produces a Coulomb potential
with a far more rapid spatial variation than that of the frequently-invoked
continuum distribution of the ''jellium'' models \cite{brack}. The
linearization in $\left( 5\right) $ can still be made under the following
assumptions. Our main observation at the outset is that in a high-density
electron gas, any electric charge is screened out very rapidly, namely, at
distances beyond a characteristic Debye screening length, say $q_{0}^{-1}$
(which is inversely proportional to $\sqrt{ k_{F}}$, as can be seen below).
Also, we notice that the slow variation of $V\left( {\bf r}\right) $ is
usually supposed to be over a de Broglie wavelength for an electron at the
Fermi surface, that is $2\pi /k_{F}$. From this view, we may say that a
possible conflict with the use of the Thomas-Fermi approximation occurs only
close to the positive ions, closer than a shielding distance $q_{0}^{-1}$.
Disregarding this limitation, we use the Thomas-Fermi approximation in $%
\left( 5\right) $ to obtain, after a straightforward integration, 
\begin{equation}
\rho \left( {\bf r}\right) =\rho _{0}-\frac{q_{0}^{2}}{4\pi }V\left( {\bf r}%
\right) \;\;\;,  \label{six}
\end{equation}
where $q_{0}^{2}=\frac{4k_{F}}{\pi a_{H}}$, with $a_{H}$ the Bohr radius.
The results are not strongly affected by this approximation. For example,
good agreement within natural limits has been obtained previously\cite{despa}
for the fullerene molecule described in this way and without the simplifying
linearization. At the same time, we can see that the density follows the
potential closely, which means that the validity of the theory is ensured,
as we already stipulated above, by an appropriate requirement on the
electron density. Of course this theoretical model loses its validity at
large distances from the ion locations because the electron density
vanishes, and at very short distances, towards the center of the cluster
cage, where the density becomes infinite with the potential \cite{note2}.

Eq. $\left( 6\right) $ is a quasi-classical result obtained in the high
density approximation for the valence electrons. Within the quasi-classical
approximation,\cite{note3} local variations of the electron density leave
the exchange contribution unchanged, as a consequence of its nonlocal
(quantum) character. Therefore, a first-order quantum correction to this
quasi-classical result represents the exchange energy 
\[
E_{ex}=-\frac{3}{4}\left( \frac{3}{\pi }\right) ^{\frac{1}{3}}\rho _{0}^{%
\frac{4}{3}}\;\;\;, 
\]
Consequently, we may assume that the common potential $V\left( {\bf r}%
\right) $ is generated only by the electron distribution in the presence of
the discrete ionic background.\ We may therefore set up the basic Poisson
equation to yield 
\begin{equation}
\Delta V=4\pi \rho _{0}-q_{0}^{2}V\left( {\bf r}\right) -4\pi
\sum_{i}^{N}z_{i}\delta \left( {\bf r-R}_{i}\right) \;\;\;.  \label{seven}
\end{equation}
We have to solve a self-consistent field problem which accounts for the
electron distribution profile in the presence of a discrete positive
background. The last term on the right side of eq.$\left( 7\right) $
represents the density of positive charge with $R_{i}$ the average distance
of an ion from the center of the cluster and $i$ is an index running over
the ions, each with electric charge $z_{i}$. These locations are chosen
without regard to the stability of the configuration.

According to the principle of superposition, Poisson's equation $\left(
7\right) $ may separate: 
\begin{equation}
\Delta \left( V_{1}+V_{2}\right) =4\pi \rho
_{0}-q_{0}^{2}V_{1}-q_{0}^{2}V_{2}-4\pi \sum_{i=1}^{N}z_{i}\delta \left( 
{\bf r-R}_{i}\right) \;\;\;,  \label{eight}
\end{equation}
which means that we have to solve two simpler equations rather than one very
complex equation. The first is given by 
\begin{equation}
\Delta V_{1}=4\pi \rho _{0}-q_{0}^{2}V_{1}\;\;\;,  \label{nine}
\end{equation}
and represents the effective electrostatic potential due to the electron
self-distribution where the discrete nature of the positive charges is
disregarded. This equation will be solved inside a large sphere of radius $R$%
, which has to contain most of the valence electron density.\cite{chelik}
The second equation becomes 
\begin{equation}
\Delta V_{2}=-q_{0}^{2}V_{2}-4\pi \sum_{i=1}^{N}z_{i}\delta \left( {\bf r-R}%
_{i}\right) \;\;\;,  \label{ten}
\end{equation}
and accounts for the remaining terms of the total potential. The discrete
nature of the positive background is employed here.

\medskip

By Fourier transformation, the latter equation becomes 
\begin{equation}
\int d{\bf k}\;V_{2}\left( k\right) \left( k^{2}-q_{0}^{2}\right) \exp
\left( i{\bf kr}\right) =\frac{1}{2\pi ^{2}}\sum_{i=1}^{N}z_{i}\int d{\bf k}%
\;\exp \left[ i{\bf k}\left( {\bf r-R}_{i}\right) \right] \;\;,  \nonumber
\end{equation}
wherefrom 
\[
V_{2}\left( k\right) =\frac{1}{2\pi ^{2}}\sum_{i=1}^{N}z_{i}\frac{\exp
\left( -i{\bf kR}_{i}\right) }{k^{2}-q_{0}^{2}}\;\;\;\;, 
\]
and the potential is simply 
\begin{equation}
V_{2}\left( {\bf r}\right) =\frac{1}{2\pi ^{2}}\sum_{i=1}^{N}z_{i}\int d{\bf %
k}\;\frac{\exp \left[ i{\bf k}\left( {\bf r-R}_{i}\right) \right] }{%
k^{2}-q_{0}^{2}}\;\;,  \label{eleven}
\end{equation}
or 
\begin{eqnarray}
V_{2}\left( {\bf r}\right) &=&\frac{2}{\pi }\sum_{i=1}^{N}z_{i}%
\sum_{lm}i^{l}\int dk\;\frac{k^{2}}{k^{2}-q_{0}^{2}}j_{l}\left( k\left| {\bf %
r-R}_{i}\right| \right)  \nonumber \\
&&\int d\Omega _{k}Y_{lm}^{*}\left( \theta _{k},\varphi _{k}\right)
Y_{lm}\left( \theta _{i},\varphi _{i}\right) \;\;,  \nonumber
\end{eqnarray}
in terms of spherical Bessel functions $j_{l}\left( k\left| {\bf r-R}%
_{i}\right| \right)$. After the integration over $\Omega _{k},$ the above
equation for the potential reduces to 
\begin{equation}
V_{2}\left( {\bf r}\right) =\frac{2}{\pi }\sum_{i=1}^{N}z_{i}\int dk\;\frac{%
k^{2}}{k^{2}-q_{0}^{2}}\sin \left( k\left| {\bf r-R}_{i}\right| \right) \;\;.
\label{twelve}
\end{equation}

\medskip The former equation $\left( 9\right) $ (subject to appropriate
boundary conditions, as we will see below) deals with the effective Coulomb
potential due to the electron distribution in the super-sphere of effective
radius $R.$ To solve it we exploit the fact that $\left( 9\right) $
separates in spherical polar coordinates $r,\;\theta ,\;\varphi $. The
solution of Poisson's equation $\left( 9\right) $ is given by 
\begin{equation}
V_{1}\left( r,\theta ,\varphi \right) =\sum_{l,m}F_{lm}\left( r\right)
Y_{lm}\left( \theta ,\varphi \right) \;\;.  \label{thirteen}
\end{equation}
Each $F_{lm}$ is actually independent of $m$ and satisfies the radial
equation 
\begin{equation}
\frac{1}{r}\frac{d^{2}}{dr^{2}}\left( rF_{l}\right) -\frac{l\left(
l+1\right) }{r^{2}}F_{l}=4\pi \rho _{0}-q_{0}^{2}F_{l}\;\;.  \label{fourteen}
\end{equation}
Hence we now drop the subscript $m$. Strictly, terms corresponding to $l=0$
have been considered separately in solving the above equation. The general
solution for the radial equation is 
\begin{eqnarray}
F_{l}\left( r\right) &=&\frac{\left( 4\pi \right) ^{3/2}}{q_{0}^{2}}\rho
_{0}+\frac{A_{00}}{r}\sin \left( q_{0}r\right) +\frac{B_{00}}{r}\cos \left(
q_{0}r\right)  \label{fiveteen} \\
&&+\sum_{j=0}^{l}\frac{C_{lj}}{q_{0}^{j}r^{j+1}}\left[ B_{lm}\exp \left(
-q_{0}r\right) +\left( -1\right) ^{j}A_{lm}\exp \left( q_{0}r\right) \right]
\;\;,  \nonumber
\end{eqnarray}
where 
\begin{equation}
C_{lj}=\frac{l\left( l+1\right) \left( l+j\right) !}{2^{j}j!\left(
l-j\right) !}\;\;\;,  \label{sixteen}
\end{equation}
and $A_{00},\;B_{00},\;A_{lm}\;$and $B_{lm}$ are constants that will be
determined. The effective Coulomb potential $V_{1}$ can be written then as 
\begin{eqnarray}
V_{1} &=&\frac{4\pi }{q_{0}^{2}}\rho _{0}+\frac{A_{00}}{r}\sin \left(
q_{0}r\right) +\frac{B_{00}}{r}\cos \left( q_{0}r\right)  \label{seventeen}
\\
&&+\sum_{lm}\;^{\prime }\sum_{j=0}^{l}\frac{C_{lj}}{q_{0}^{j}r^{j+1}}\left[
B_{lm}\exp \left( -q_{0}r\right) +\left( -1\right) ^{j}A_{lm}\exp \left(
q_{0}r\right) \right] Y_{lm}\left( \theta ,\varphi \right) \;\;,  \nonumber
\end{eqnarray}
where the prime in the right hand term of the equation means that the
summation over $l$ begins from $l=1$. This potential has to be finite for $%
r=0,$ which means that $B_{00}=0,\;$and $B_{lm}=(-1)^{l+1}A_{lm}$.

Taking $\left( 12\right) $ into account, we find the total effective Coulomb
potential inside the super-sphere is 
\begin{eqnarray}
V_{in} &=&\frac{4\pi }{q_{0}^{2}}\rho _{0}+\frac{A_{00}}{r}\sin \left(
q_{0}r\right) +V_{2}\left( {\bf r}\right)  \label{eighteen} \\
&&+\sum_{lm}\;^{\prime }\sum_{j=0}^{l}\frac{C_{lj}\left( -1\right) ^{j}}{%
q_{0}^{j}r^{j+1}}A_{lm}\left[ \exp \left( q_{0}r\right) +\left( -1\right)
^{l-j}\exp \left( -q_{0}r\right) \right] Y_{lm}\left( \theta ,\varphi
\right) \;\;,  \nonumber
\end{eqnarray}
everywhere except for ${\bf r}={\bf R}_{i}$. Outside the super-sphere, a
Laplace equation applies and the solution vanishing at infinity is 
\begin{equation}
V_{out}=\frac{B_{00}}{r}+\sum_{lm}\;^{\prime }\frac{B_{lm}}{r^{l+1}}%
Y_{lm}\left( \theta ,\varphi \right) \;\;.  \label{nineteen}
\end{equation}
If the potential is specified on the surface of the bounding sphere, the
coefficients entering $\left( 17\right) $ and $\left( 18\right) $ can be
determined by evaluating $V\left( R,\theta ,\varphi \right) $ and using 
\begin{equation}
A_{lm}=\int \;d\Omega \;Y_{lm}^{*}\left( \theta ,\varphi \right) g\left(
\theta ,\varphi \right) \;\;,  \label{twenty}
\end{equation}
where $g\left( \theta ,\varphi \right) $ is an arbitrary function. Here, $g$
represents a ''pseudo-charge density'' designed to be a smooth, nodeless
function which, in order to maintain the electrical neutrality of the entire
system, has to agree exactly with the true charge density outside the region
bounded by the super-sphere of radius $R$.

An additional comment is appropriate here regarding the present theory. We
begin by asking,``How unique is the potential in eqs. $\left( 18\right) $
and $\left( 19\right) $?" If we demand that our ''pseudo-charge density'' $g$
agrees with the true charge density outside the ''super-sphere'' then the
potential is uniquely determined. The inside region is not uniquely fixed by
this procedure; however, if we require that the total charge of the valence
electrons be normalized, then the fraction of electronic charge contained in
this region must be large, e.g., more than $\sim 95\%$ of the total. This
means that the behavior of the charge contained in this region must dominate
the static properties of the metallic cluster.

The effective cluster potential given by $\left( 18\right) $ displays the
usual collective aspects of the electron gas. The primary manifestations of
the collective behavior are a) collective oscillations of the valence
electrons as an entity, the so-called ''plasma'' oscillations, and b) the
screening of the field of any individual electric charge beyond a
characteristic length $q_{0}^{-1}.$ The former is fundamentally a
diffraction effect, the electron wave nature being essentially disregarded
in this kind of calculation. The screening of the ionic fields causes the
remainder of the electron gas to stay diffuse, and so leads to a deficiency
of negative charge just outside the immediate neighborhood of each positive
ion enclosed in its neutralizing, co-moving electronic cloud. Thus, the
cluster potential exhibits additional spatial oscillations which are not
determined solely by its behavior in the neighborhood of ${\bf r}$. In a
collective oscillation, each individual electron suffers a small periodic
perturbation of its velocity (recall that the electron density $\left(
5\right) $ is a result of the perturbation of the kinetic operator) and
position due to the combined potential of all the other particles, both
positive and negative. The cumulative potential of all the electrons may be
quite large since the long range of the Coulomb interaction permits all the
electrons to contribute to the potential at every point. The collective
behavior of the electron gas dominates phenomena involving distances greater
than the characteristic length $q_{0}^{-1},$ while the individual particle
component is associated with the random thermal motion of the electrons. In
the approximate level of this analysis, the effects of collective excitation
on the correlation are neglected, as a second-order effect.

Usually, the long range of the Coulomb interactions having the character of $%
\left( 18\right) $ precludes immediate application of these results to the
calculation of the ground-state energy of the cluster. Therefore, we are not
able to perform a minimization of the ground-state energy with respect to
the volume of the super-sphere. Consequently, the self-consistency of the
potential is affected by this lack of information.

With all the assumptions of the model and its mathematical output now
presented, we may already point out some general characteristics we may
expect for the behavior of the effective cluster potential. Since in the
present perturbation approach the expansion of the electron density is based
on plane waves, the cluster potential displays a high value in the central
region. The potential is strongly dependent on the Coulomb correlations and,
naturally enough, very sensitive to the position of the positive ions. (We
discuss this aspect later.) Nonlocal effects due to the particle spins in
the mean field for electrons are disregarded. Hence the method produces
state-independent potentials.

In the following discussion, we work out an example of a metallic cluster $%
M_{13}$ with icosahedral symmetry that closely approximates spherical
symmetry. The model for the ionic cores is that of hard-spheres occupying a
total volume in space equal to $\Omega _{ions}$. The valence electrons are
highly confined between the ionic cores. The unperturbed density $\rho _{0}$
is expressed by 
\begin{equation}
\rho _{0}=\frac{3}{4\pi r_{s}}=\frac{N}{\frac{4\pi }{3}R^{3}-\Omega _{ions}}%
\;\;,  \label{twenty-one}
\end{equation}
where $r_{s}$ is a point in the space available to the electrons, the
``electronic interspace", outside the ion cores. This means that we have
subtracted from the entire volume of the super-sphere of radius $R$ the
volume assigned to the ionic cores $\Omega _{ions}$; $N$ is the total number
of the delocalized electrons, equal to the number of ionic charges. The
distance between the centers of the central and outer ions is the bond
length. For numerical calculation we set $R_{i}=5.9\;a.u.$. The core volume $%
\Omega _{ions}$ is usually computed by taking into account the ionic radius;
In our first example, we let this radius be $2.74 a.u.$ so the resulting
volume $\Omega _{ions}$ is $1122 (a.u.)^{3}$. For the other two cases, this
volume is a free parameter. The electronic interspace was chosen to be $%
r_{s}=0.75\;a.u.$, in accordance with the high-density electron gas
requirement $\left( r_{s}\ll 1\right) $, and by imposing that $95\%$ of the
total electrons must be inside the super-sphere, the super-sphere radius
becomes $R=6.5\;a.u.$ In Fig. 1 we can see the corresponding effective
potential inside the super-sphere as a function of $r$ and $\theta .$ In
Fig. 2, the spatial dependence of the same potential is displayed along the
coordinates $\theta $ and $\phi $ at the radius where the outer ions lie, $%
R_{i}=5.9\;a.u.$. These pictures show the regular, collective
characteristics we discussed above. The oscillations we observe are a
manifestation of the self-consistency of Poisson's equation and represents
the main correlation effect of the electron gas in the metallic state (the
high-density limit).

Despite the collective aspects which contribute to the mean-field character
of the effective potential, $V\left( {\bf r}\right) $ remains sensitive to
the geometry and composition of the cluster. If, for example, one host atom
in the cluster cage is replaced by an impurity atom $A$, the potential
reflects this structural change. We explore this property in the following.
Let us assume that the impurity $A$ is a trivalent metal atom $\left(
A\right) $ which releases all three of its valence electrons into the Fermi
sea in the bulk volume. We take its core volume to be the same as the $M$
ions. The substitution does not change the symmetry and we assume that the
most stable geometry of the cluster has the dopant at the centre. Figs. 3
and 4 show the corresponding effective electrostatic potential for the $%
AM_{12}$ system. By comparing this with the effective mean-field potential
for $M_{13}$, we observe that the presence of the trivalent atom at the
center makes the potential much deeper. (See Fig. 3, where $V$ is displayed
as a function of $r$ and $\theta $ .). The delocalized electrons polarize
inward, toward the high Coulomb field of the central, trivalent ion. Fig. 4
shows also that the shape of the quantum oscillations at the cluster surface
changes relative to the previous case. The amplitudes of oscillations become
rather uniform which means that the screening among the electric charges is
much better for this system than for the homogeneous cluster, a consequence
of the increased number of delocalized electrons, from $13$ to $15$.
Moreover the better screening effect here results in a change in the
effective force acting in the electron gas. This can be seen in Fig. 4 as a
phase-shift of electron density oscillations at the position of the positive
ions. (Compare this with the effective potential displayed in Fig. 2.)

If we move the trivalent ion to the outer shell, the disturbance of course
goes toward the surface (see Fig. 5) and the electron density is enhanced
around the vertex where the trivalent ion is located. This behavior of the
effective potential is supplemented by the appearance of more pronounced
Coulombic correlations of the valence electrons near the surface. The
quantum oscillations are sensibly disturbed by the trivalent impurity
located on the cluster surface. This disturbance appears as irregular
behavior along the $\theta $ coordinate at constant $r$. Also, a large
potential difference, about $1\;a.u.$, can be seen in Fig. 6 between the
position of the trivalent impurity $\left( \theta =0,\;\varphi =0\right) $
and the antipodal position $\left( \theta =\pi ,\;\varphi =0\right) $
occupied by a host ion. The potential difference leads to a displacement of
the electronic cloud towards the position of the trivalent ion and a
deficiency of negative charge in the opposite direction. Consequently, a
diffusive trend of electron density oscillations can be observed in the
hemisphere at $\theta =\pi $.

We may conclude that the effective electrostatic potentials for metallic
clusters is subject to important Coulombic correlation effects which can be
visualized at the proper scale by employing a discrete description for the
positive background. The electron density shows a static screening which is
rather localized near the positive charges and supplemented by the
long-range oscillatory behavior. For singly-doped binary metallic systems,
the depth of the effective electrostatic potential depends on the dopant
position in the cluster geometry. The collective aspects of the excitations
of electrons delocalized through the cluster volume are strongly perturbed
by the presence of the impurity. The main effect of these structural
rearrangements of the ions is the change of the effective potential, as we
have shown. The change of the cluster potential will alter, in turn, the
ordering of the related electron shell, a fact which has been observed in
many experiments \cite{heinebrodt,class,akola}. Therefore our findings may
be interpreted as qualitative support for various models explaining the
shell inversions for doped metallic clusters \cite
{kappes,yeretzian,baladron,yannouleas}. Obviously other kinds of changes of
dopant atoms may induce still different effects, that will depend on their
locations in the cluster.

Finally, we may say that the method developed here is simple and flexible
and can yield, to some extent, accurate approximations to the exact
effective potentials with minor computing effort. Also, it has the advantage
of physical immediacy, i.e., the present approach is easy to interpret. This
makes the method useful for a fast check of the effective potential to
systems, clusters of heavy elements, for example, presently beyond the
capability of more accurate approaches.

\vskip 1em

{\bf Acknowledgement}

This research was supported by a Grant from the National Science Foundation.

\begin{center}
\newpage Figure captions
\end{center}

Fig. 1

The effective electrostatic potential inside the cluster cage for $M_{13}$
for $0<r<6.5\;a.u.,$ $0<\theta <\pi $ $rad$ and $\varphi =0.$

Fig. 2

The same effective electrostatic potential displayed as a $\theta -\varphi $
plot at the position of the surface ions.

Fig. 3

The corresponding effective electrostatic potential for $AM_{12}$ system for 
$0<r<6.5\;a.u.,$ $0<\theta <\pi \;rad$ and $\varphi =0$ with the impurity in
the center of the icosahedral cluster cage.

Fig. 4

The effective electrostatic potential showed in Fig. 3 displayed here along
the coordinates $\theta $ and $\varphi $ at the position of surface ions.

Fig. 5

A plot analogous to that of Fig.3, for the $AM_{12}$ system with impurity at
the vertex, $\left( \theta =0,\varphi =0\right) $ and for $0<r<6.5\;a.u.$.

Fig. 6

The $\theta -\varphi $ spatial dependence of the potential displayed in Fig.
5 at the position of the outermost ion shell. The electrons must be in their
ground state and highly confined in the cluster cage.


\begin{references}
\bibitem{note2}  To remove this qualitative defect, we should work with the
electron density within primary form $\left( 5\right) $, which remains
finite at the origin for a potential which is singular as $r^{-1}$. Further
employment of eq $\left( 6\right) $ for the electron density has the
advantage of producing an analytical result for the effective cluster
potential.
\end{references}
\end{document}